%Paper: hep-th/9501118
%From: morchio@morchio.difi.unipi.it
%Date: Thu, 26 Jan 95 09:59:25 EST

\magnification 1200

%def. matem.

\def\reali{\hbox{\rm I\hskip-2pt\bf R}}
\def\realipunto{\hbox{\rm I}\hskip-2pt \dot {\bf R}}

\def\interi{\hbox{\rm{Z\hskip-8.2pt Z}}}
\def\naturali{\hbox{\rm I\hskip-5.5pt\bf N}}

%spaziature

\def\bsk{\bigskip}
\def\msk{\medskip}
\def\ssk{\smallskip}
\def\ni{\noindent}

% Fonti di caratteri

\font\titfnt=cmbx10 at 14.40 truept

% Definizione formato della pagina

\newdimen\pagewidth \newdimen\pageheight \newdimen\ruleht
\hsize=31pc  \vsize=45pc  \maxdepth=2.2pt  \parindent=19pt
\pagewidth=\hsize \pageheight=\vsize \ruleht=.5pt
\abovedisplayskip=10pt minus 2pt
\belowdisplayskip=11pt minus 2pt
\abovedisplayshortskip=8pt minus 2pt
\belowdisplayshortskip=9pt minus 2pt

\baselineskip=15pt minus 0.5pt
\lineskip=2pt minus 0.5pt
\lineskiplimit= 2pt

\mathsurround=1pt

% Definizioni per l'output

%(numeri in alto alternati)
\nopagenumbers
\def\testos{\null}
\def\testod{\null}
\headline={\if T\tpage{\gdef\tpage{F}{\hfil}}
            \else{\ifodd\pageno\rightheadline\else\leftheadline\fi}
           \fi}
\gdef\tpage{T}
\def\rightheadline{\hfil{\tensl\testod}\hfil{\tenrm\folio}}
\def\leftheadline{{\tenrm\folio}\hfil{\tensl\testos}\hfil}

%(numeri in alto al centro)
%\nopagenumbers
%\headline={\hss\tenrm\folio\hss}
%\voffset=2\baselineskip

% Registrazione referenze su file esterno e macro di conteggio

\newcount\numref
\global\numref=1
\newwrite\fileref
\immediate\openout\fileref=ref.tmp
\immediate\write\fileref{\parindent 30pt}
\def\citaref#1{${[\the\numref]}$\immediate\write\fileref
              {\par\noexpand\item{{[\the\numref]\enspace}}}\ignorespaces
              \immediate\write\fileref{{#1}}\ignorespaces
              \global\advance\numref by 1\ignorespaces}

%Fine delle def. generali

\def \H {{\cal H}}
\def \vp {{\varphi}}
\def \A {{\cal A}}
\def \D {{\cal D}}
\def \F {{\cal F}}
\def \Z {{\cal Z}}
\def \Ldue {L^2 ([0, 2\pi) \times \reali, \, d\varphi \, dA )}
\def \LdueQth {L^2 (\reali \times [0,2\pi), \, dQ \, d\theta )}
\def \LdueQq {L^2 (\reali \times [0,2\pi), \, dQ \, dq )}
\def \thm {{\theta_M}}
\def \undp {{1\over 2 \pi}}
\def \thf {{\theta_F}}
\def \unm {{1\over 2}}
\def \cs {C^{*}}
\def \dm {d\mu}
\def \dmT {d\mu_T}
\def \dmz {d\mu^0}
\def \xA {{ (x(\tau) , A(\tau)) }}
\def \vpA {{ (\vp(\tau) , A(\tau)) }}
\def \xt {{ (x(\tau)) }}
\def \vpt {{ (\vp(\tau)) }}
\def \At {{ (A(\tau)) }}
\def \piz {{\pi_0}}
\def \Hth {{{\cal H}_\theta}}
\def \Aobs {{\cal A}_{obs}}
\def \Fobs {{\cal F}_{obs}}
\def \intinf {\int_{-\infty}^{+\infty}}
\def \suminf {\sum_{n=-\infty}^{\infty}}
\def \xm {{x_{-}}}
\def \xp {{x_{+}}}
\def \vpm {{\vp_{-}}}
\def \vpp {{\vp_{+}}}
\def \Am {{A_{-}}}
\def \Ap {{A_{+}}}
\def \DA {\D A}
\def \Dx {\D x}
\def \psith {{\psi_\theta}}
\def \psizth {{\psi_\theta^0}}
\def \AQP {{\cal A}_{QP}}
\def \Aqp {{\cal A}_{qp}}
\def \endproof {$q.e.d.$}
\def \intdp {{\int_{0}^{2 \pi}}}
\def \intinf {{\int_{-\infty}^{\infty}}}
\def \intT {{\int_{-T}^{T}}}
\def \intt {{\int_{-\tau}^\tau}}
\def \KfvpA {K_\tau^\thf(\vp , A , \vp ' , A ')}
\def \KfvpAdue {K_{2 \tau}^\thf(\vp , A , \vp ' , A ')}
\def \ittH {\exp - \tau H}

%Fine delle definizioni

\hfill  IFUP--TH  7/95
\msk
\centerline{\titfnt The $(QED)_{0+1}$ model}
\ssk
\centerline{\titfnt and a possible dynamical solution}
\ssk
\centerline{\titfnt of the strong $CP$ problem}

\bsk
\centerline{J. L\"offelholz$ ^{*}$ }

\centerline{\it Mathematisches Institut,
Universit\"at Leipzig, Germany}

\msk
\centerline{G. Morchio}

\centerline{\it Dipartimento di Fisica dell'Universit\`a and INFN,
           Pisa, Italy}

\msk
\centerline{F. Strocchi}

\centerline{\it Scuola Normale Superiore and INFN,
              Pisa, Italy}

\bsk\msk
\ni {\bf Abstract}.
The $(QED)_{0+1}$ model describing a quantum mechanical particle
on a circle with minimal electromagnetic interaction and with a
potential $-M \cos(\vp -\thm)$, so that it mimics the massive Schwinger
model, is discussed as a prototype of mechanisms and infrared structures
which characterize gauge quantum field theories in positive gauges
and $QCD$ in particular.
The functional integral representation in terms of the field
variables which enter in the Lagrangean displays non--standard
features, like a complex functional measure (failure
of Nelson positivity), a crucial r\^ole of the boundary conditions,
and the decomposition into $\theta$ sectors already in finite volume.
In the infinite volume limit, one essentially recovers the standard
picture when $M=0$ (\lq\lq massless fermions\rq\rq ), but one meets
substantial differences for $M\neq 0$: for generic boundary conditions,
independently of the lagrangean angle of the topological term,
the infinite volume limit selects the sector with $\theta = \thm$
and provides a natural \lq\lq dynamical\rq\rq\ solution of
the strong $CP$ problem.
In comparison with previous approaches, the strategy discussed here
allows to exploit the consequences of the $\theta$ dependence
of the free energy density, with a unique minimum at $\theta = \thm$

\msk
\ni $ ^{*} $ Supported by
Deutsche Forschungsgemeinschaft,
DFG Nr. Al 374/1-2

\vfill
\eject

\ni
{\bf 1. Introduction.}

\bsk
The use of functional integral methods has lead to the
discovery \citaref{G. 't Hooft, Phys.  Rev. Lett.,
{\bf 37}, 8 (1976); R. Jackiw and C. Rebbi, Phys.  Rev. Lett. {\bf 37},
172 (1976); C.G. Callan, R.F. Dashen and D. Gross, Phys.  Lett. {\bf
B63}, 34 (1976).} \citaref{J. Fr\"{o}hlich, in {\it Renormalization
Theory}, G. Velo and A.S. Wightman eds., D. Reidel 1976 and references
therein; E. Seiler, {\it Gauge Theories as a Problem of Constructive
Field Theory and Statistical Mechanics},
Lecture Notes in Physics vol.159, Spinger--Verlag 1982. }
\citaref{For a general account see R.
Jackiw's  contribution to S.B. Treiman, R. Jackiw, B. Zumino and E.
Witten, {\it Current Algebra and Anomalies}, World Scientific 1985.}
 of important non--perturbative features
of gauge quantum field theories (QFT), in particular the mechanism
of $\theta$--vacua, the winding number picture, the $U(1)$
chiral symmetry breaking, the topological terms, the strong $CP$
problem etc. A rigorous analysis of the massive Schwinger
model reproducing such features [2] has further backed the standard
wisdom on QCD [3].

The problems arise when such picture is confronted with the standard
perturbative approach, where it seems difficult to incorporate
a non vanishing order parameter $\bar q q$
\citaref{R.J. Crewther, in {\it Field Theoretical Methods in Particle
Physics}, W. R\"{u}hl ed., D. Reidel 1980.}
and, in the presence of
a fermion mass term,  a \lq\lq natural\rq\rq\
strong $CP$ symmetry
\citaref{S. Weinberg, Phys.
Rev.  Lett. {\bf 40}, 233 (1978); F. Wilczek, ibid 279 (1978);
R.D. Peccei and H.R. Quinn, Phys. Rev.  Lett. {\bf 38}, 1440 (1977).},
and the mechanism suggested to solve
such problems (non--integer winding numbers, Goldstone dipole,
Peccei--Quinn symmetry) are not without difficulties.

The aim of the present note is to revisit the above problems
on the basis of a rigorous functional integral analysis of
a simple model (which mimics, in $0+1$ dimensions, the massive
Schwinger model): as we shall see the $CP$ conserving condition,
$\theta = \theta_M$, ($\thm$ the fermion mass angle) will emerge
as a dynamical effect in the thermodynamical limit of the
functional integral, generically in the boundary conditions,
with no need of fine tuning.

The model ($QED_{0+1}$)
describes a quantum mechanical particle on a circle
with minimal electromagnetic interaction and with a potential
$-M \cos(\vp -\thm)$, which mimics the fermion mass term
in the massive Schwinger model;
the coordinate $\vp$ is the analog of the scalar field which bosonizes
the fermions in $1+1$ dimensions, and that is why it lives
on a circle
\citaref{F. Acerbi, G. Morchio and F. Strocchi, J. Math.
Phys. {\bf 34}, 899 (1993);
Lett.  Math.  Phys. {\bf 27}, 1 (1993).
For a different discussion of the $(QED)_{0+1}$ model see
J. L\"{o}ffelholz, Helv.  Phys.  Acta {\bf 64}, 484 (1991).
See also K. Skenderis and P. Van Nieuwenheuzen,
Stony Brook preprint 1994.}.
Without loss of generality, $M$ can be taken non--negative.
The model is easily analysed in the
Hamiltonian approach,
$$ H = {1 \over 2} (p - eA)^2 + \unm E^2 - M \cos(\vp - \thm)
  \eqno(1) $$
($E = \dot A$, $p = \dot \vp + eA$), in terms of the canonical
\lq\lq field\rq\rq\ $\cs$algebra $\A$ generated by
$\exp i \varphi$,
$\exp i \alpha A$,
$\exp i\beta p$, $\exp i \gamma E$,
$\alpha , \beta , \gamma  \in \reali$.
For $M > 0$ and small, a unique ground state exists,
(its energy has an essential singularity at $e=0$),
it yields a
non--regular (i.e. non Schroedinger) representation of $\A$,
and a reducible representation of the gauge invariant
observable subalgebra $\Aobs$ generated by
$\exp i \vp$, $\exp i \beta (p-eA)$, $\exp i \gamma E$,
$\beta, \gamma \in \reali$. The irreducible representations
of $\Aobs$ are labelled by an angle $\theta$;
the unique ground state of $H$
belongs to the sector $\theta =\thm$ and is invariant under the
$CP$ symmetry ($\vp \mapsto -\vp -2\thm$ mod $2\pi$, $A \mapsto -A$).

The so obtained naturality of the $CP$ conserving condition,
$\theta =\thm$, crucially depends on the strategy (also followed
by the standard perturbative approach) of formulating the model
in terms of a {\it field} algebra; the alternative strategy
which restricts the attention to the {\it observable} algebra and
its irreducible representations does not allow for a dynamical
choice between the various $\theta$ sectors, each with a lowest energy
state, $\psi^0_\theta$; each value of $\theta$ is then allowed
and the $CP$ conserving condition becomes accidental; on the other hand,
such an approach is intrinsically non--perturbative and
the naturality condition (i.e. stability under higher order
perturbative corrections)
\citaref{S. Weinberg, Phys.  Rev. Lett. {\bf 29}, 388
(1972); Rev. Mod.
Phys. {\bf 46}, 255 (1974); H. Georgi and S.L.
Glashow, Phys.  Rev. {\bf D6}, 2977 (1972).}
cannot even be posed.

The non--trivial information that among the $\theta$ sectors the
minimum of the energy is reached for $\theta =\thm$ (provided
by the first approach) has important consequences also for
the functional integral approach, which is done in terms
of lagrangean variables.
As a matter of fact, the functional integral approach to the model
is very instructive since it displays general features and
non--standard mathematical properties which are likely to be shared
by the functional integral approach to four--dimensional
gauge theories, in positive gauges.

The first lesson is that the naive (but popular) euclidean
functional integral representation in infinite volume (here
infinite time)
$$ d\mu \, \vpA = \Dx \; \DA \; e^{ - \int (\unm  \dot \vp^2
+ \unm \dot A^2 + i e \dot \vp A - M \cos (\vp - \thm) +
i \theta_L  \dot A ) \, d\tau }  \equiv  $$
$$ \equiv  d\mu^{free} \, \vpt \; d\mu^{free} \, \At
    \; e^{ \int ( - i e \dot \vp A
     - i \theta_L \dot A  + M \cos (\vp - \thm)) \, d\tau } \eqno(2) $$
(where to be general a topological term, with lagrangean parameter
$\theta_L$  has been included)
is highly misleading, if not wrong. One of the reasons is that
it involves infrared singular variables (here the variable $A$):
even in the presence of an ultraviolet cutoff,
the ground state defines a non--regular representation of the CCR
algebra generated by those variables, and
a change of representation takes then place in the infinite volume
limit ({\it infrared renormalization})
\citaref{J. L\"{o}ffelholz, G. Morchio and F.
Strocchi, Spectral stochastic processes arising in quantum mechanical
models with a non--$L^2$ ground state, preprint IFUP-TH 55/94,
Lett. Math. Phys., in press}.

One can show that in finite \lq\lq volume\rq\rq , $\tau \in [-T,T]$,
eq.(2) gives rise to a well--defined measure $\dmT$, and in
particular there are no ultraviolet problems for the term $\dot \vp A$,
contrary to the case of a particle in a magnetic field (here
$A = A(\tau)$, rather than $A(x(\tau))$.
However, the control of the infinite volume limit $T \to \infty$
exhibits completely new features with respect to ordinary
quantum mechanical models, like non--relativistic
particles with velocity--independent potentials
and scalar field models:

\ssk\ni
i) $\dmT$ is  {\it complex};
Nelson positivity does not hold, so that
the continuity of the functional (on the space of continuous
functions of the trajectories)
which defines $\dmT$ (by the Riesz--Markov theorem)
is not automatic.
Such continuity, which is necessary (and sufficient)
for $\dmT$ to be a measure, rather than merely a
\lq\lq cylinder measure\rq\rq , only holds in finite volume.
In the infinite volume limit, the correlation functions
have a functional integral representation only in terms of
a complex cylinder measure, with infinite total variation.

\ssk\ni
ii) a crucial r\^ole is played by the Osterwalder--Schrader
(OS) positivity condition, also in connection with the
problem of {\it reducibility} on the observable algebra of
the functional defined by $\dmT$, which gives the
$\theta$ {\it angle structure}.

\ssk\ni
iii)  an important r\^ole is played by the boundary conditions, which
may now be complex and are only constrained by OS positivity;
they provide the correct way to achieve the reduction into
$\theta$ sectors, which occurs already {\it in finite volume},
in terms of functional integrals with winding numbers $n$ and
phases $\exp i n \theta$ (the popular claim that this is obtained
by the gauge invariance condition is misleading, see below and
\citaref{J. L\"{o}ffelholz, G. Morchio and F. Strocchi,
in preparation.}).
However, the winding number interpretation loses its meaning
in the infinite volume limit.

\ssk
In agreement with the results of the Hamiltonian approach,
the infinite volume limit, $T \to \infty$,
of the correlation functions of the
observables is strongly affected by the dependence of
the \lq\lq free energy density\rq\rq\ on the $\theta$ parameter:

\ni
a) for $M \neq 0$, {\it generically in the boundary conditions},
the limit $T \to  \infty$ gives the correlation functions on
the ground state, which belongs to
the sector with $\theta =\thm $. This gives
a {\it dynamical solution of the strong $CP$ problem}.
The results of 't Hooft analysis [1],
\citaref{ G. 't Hooft, Physics Reports, {\bf 142},
357 (1986) and references therein.}
can be obtained only
by a \lq\lq fine tuned\rq\rq\ choice of the boundary conditions,
i.e. by taking an ergodic mean on the boundary variable $A$
with weight $\exp i \theta A$; such a choice (\lq\lq non--local
in $A$\rq\rq ) is unnatural from a perturbative point of view
and it is {\it not} required by gauge invariance (see also [9]).

\ni
b) for $M=0$ one recovers the standard picture [1--3]: all the
lowest energy states in the various $\theta $ sectors
have the same energy (vacuum degeneracy); chiral symmetry is
{\it unbroken}
in the representation of the field algebra, but it is {\it broken}
in each $\theta $ sector, i.e. in each
irreducible representation of $\Aobs$;
all such representations are mathematically inequivalent but
physically equivalent, since they are related by automorphisms of
$\Aobs$ which commute with the dynamics.
Thus, the limits $M \to 0$ and $T \to \infty$ {\it do not commute},
even if the mass perturbation is well--defined and small
in each $\theta$ sector; the reduction into $\theta$ sectors only
takes place in the $M=0$ case, and its  extrapolation to
$M\neq 0$ (as implied in the standard picture [1]) is not correct.

\ssk
In conclusion, for $M=0$ the picture displayed
by the model coincides
with the standard wisdom [1--3] and the rigorous analysis
of [2] and
\citaref{J.H. Lowenstein and J.A.
Swieca, Ann. Phys. {\bf 68}, 172 (1971);
A.K. Raina and G. Wanders, ibid,
{\bf 132}, 404 (1981); N.K. Nielsen and B. Schroer, Nucl.  Phys.
{\bf B127}, 493 (1977); G.Morchio, D.Pierotti and F.Strocchi,
Ann. Phys. {\bf 188}, 217 (1988)},
whereas for $M\neq 0$ it sheads light on
the substantial differences which characterize the case of massive
fermions. In particular, the model suggests that a
\lq\lq dynamical solution\lq\lq\ of the strong $CP$ problem
takes place in the infinite volume limit, in agreement with
the arguments presented in
for the massive Schwinger model and the QCD case [13].

\bsk\bsk\goodbreak\ni
{\bf 2. Hamiltonian approach}

\bsk
The model is defined by the \lq\lq field\rq\rq\ algebra $\A$,
which can be taken as the $C^{*}$ algebra generated by
$\exp i \varphi$,
$\exp i \alpha A$,
$\exp i\beta p$, $\exp i \gamma E$,
$\alpha , \beta , \gamma  \in \reali$.
As it is typical of field
algebra arising from fermion bosonization [6], $\A$ has a
non--trivial centre, $\Z_F$, generated by $\exp 2 \pi i p$
and therefore, in each irreducible representation of $\A$,
$\exp 2 \pi i p$ is a complex number,
$\exp 2 \pi i \theta_F$, $\theta_F \in [0,1)$.
$\theta_F$ is a physically unobservable parameter (it plays the r\^ole
of the second angle in the Schwinger model [11]), since
different values of $\theta_F$ are related by the {\it gauge}
automorphisms of $\A$:
$$ \varphi \mapsto \varphi \ \ , \ \ \ \
p \mapsto p + \lambda \ \ , \ \ \ \
   A \mapsto A + \lambda / e  \ \ ,
   \ \ \ \  E \mapsto E  \eqno(3) $$
The gauge invariant observable subalgebra $\Aobs$ has a
non--trivial centre $\Z$ generated by
$\exp i q \equiv \exp i(\varphi -E/e)$,
and therefore each irreducible representation of $\Aobs$ is
labelled by the angle $\theta$ ($\theta$ {\it sector}),
defined by the value $\exp i \theta $ taken by $\exp iq$.

The Hamiltonian  $H$ takes a simple form in terms of the new canonical
variables $ Q \equiv E/e$, $P \equiv p-eA$,
$q \equiv \varphi - E/e$, $p$:
$$ H= {1 \over 2} (P^{2}+ e^{2} Q^{2}) - M \cos ( Q + q  - \theta_M )
           \eqno(4) $$
In each $\theta$ sector, $q$ can be replaced by $\theta$
and the corresponding Hamiltonian, which depends only on
$\theta - \thm $, will be denoted by
$H_\theta $.

The irreducible regular (i.e. Schroedinger) representations of
$\A $
are defined in $\Ldue$,
where $p$ acts as $ -i \partial / \partial\varphi$ with
boundary conditions
$\psi(2 \pi , A) = \psi (0,A) \exp 2\pi i \theta_F \, $.
Since $H$ is invariant under the
gauge transformations (3), its
spectrum is independent of $\theta_F$.

For $M=0$, in  $\Ldue$ the Hamiltonian (1) has only discrete
eigenvalues $E_k$, $k \in  \naturali$, with infinite multiplicity,
and the eigenvectors can be labelled by the eigenvalues
$n \in \interi$ of $p$.
The lowest energy eigenvectors $\psi^0_n$, corresponding to
$k=0$, are the strict analogues of the $n$--vacua
in the standard picture
of the massless Schwinger model [1--3][11];
they are invariant under {\it chiral transformations}, defined by
$$ \varphi \mapsto \varphi + \mu \ {\rm mod} \ 2\pi , \ \ \ \
p \mapsto p  \ \ , \ \ \ \
   A \mapsto A   \ \ ,
   \ \ \ \  E \mapsto E  \eqno(5) $$
which leave $H$ invariant. The vectors $\psi^0_n$
define reducible representations of $\Aobs$, with an
integral decomposition over the $\theta$ angle
$$ \psi^0_n =
\int_0^{2\pi} e^{i n \theta} \, \psi^0_\theta \; d\theta $$
The $\theta$--vacua $\psi^0_\theta$ do not belong to
$\Ldue$ (they define non--regular representations of $\A$, i.e.
not continuous in $\beta,\gamma,\delta$, see below).
In each $\theta$ sector {\it chiral symmetry is spontaneously
broken}.[1--3][11].

The situation is more intriguing for $M\neq 0$, since $H$ has
a purely continuous spectrum in $\Ldue$, so that there is no
ground state.

\bsk\goodbreak\ni
{\bf Theorem 1.} {\it Let $\theta _{F}\in [0 ,2\pi )$ be fixed.

\ni
i) There is a unique irreducible
representation $\piz$ of the field algebra $\A$
such that the Hamiltonian $H$,
for $M \not= 0$, is well defined and has a ground state.
Such a representation is the only one in which the spectrum of
$\exp iq$ is a pure point spectrum.

\ni
ii) The Hilbert space $\H$ of $\piz$ is given by the
Gelfand--Naimark--Segal (GNS) construction on the state
$$ \Omega_{\theta , \theta_{F}}
  (e^{i nq}  e^{i \beta p} e^{i \gamma Q + \delta P)})
=   \cases {  e^{i(n \theta +m \theta_{F})}
e^{-(e \gamma^2+ \delta^2/e)/4}
  \ \ \  & if $ \beta/ 2 \pi = m \in \interi$   \cr
  0 \ \ \ \ \ \ \ \ & otherwise. \cr }  \eqno(6) $$

\ni
iii) $\H$ can be decomposed as a direct sum of sectors
$$ \H = \sum_{\theta \in [0,2\pi)} \oplus \; \H_\theta \eqno(7) $$
The observable algebra ${\cal A}_{obs}$ acts irreducibly
in each $\Hth$. The Hamiltonian leaves each $\Hth$ invariant
and has in each $\Hth$ a pure point spectrum, with
no degeneracy. The lowest energy state in $\H$ is unique
for $M > 0$ and small, and it belongs to $\H_{\theta = \thm}$  }

\bsk\goodbreak\ni
{\bf Proof.} The Fock property of $\piz$ on the subalgebra
${\cal A}_{QP}$ generated by $ \exp i(\gamma Q + \delta P)$
is required by the existence of $ P^{2}+ e^2 Q^{2}$,
since $M \cos (Q+q-\thm)$ is always a bounded perturbation.
Furthermore, if we decompose
${\cal H}$ over the spectrum $\exp i \theta$ of $\exp iq$,
it follows that the infimum of the spectrum of $H_\theta$,
$E^0(\theta)$, has only one minimum, at $\theta = \thm$,
and therefore the existence of a ground
state implies the existence of the discrete component $\H_\thm$,
the irreducibility of the representation then implies
the discreteness of the spectrum of
$\exp i q$, i.e. equation (7).
Any vector  $\psith \in \Hth$
defines, by the Fock property of $\piz$,
a Fock state on ${\cal A}_{QP}$, and therefore,
by applying the algebra ${\cal A}_{QP}$ and taking
strong limits one can construct a vector $\psizth$ which
is a Fock no--particle state for $\AQP$.
Now, eq.(6) follows because $\psizth$ is an eigenvector
of $\exp i n q$ and $\exp 2 \pi i m p$ with eigenvelues
$\exp i n \theta$, $\exp i m \thf $, respectively;
$e^{i\beta p}$, $\beta /2\pi \notin \interi $,
changes the eigenvalue of $\exp i q$, and therefore
$e^{i\beta p} \psizth$ is orthogonal to all vectors of the
form $B \psizth$, $B\in {\cal A}_{QP}$.

In each sector $\Hth$, the state
with lowest energy is unique by a Perron-Frobenius
argument, since the kernel of $\exp -\tau H_\theta $
in the variable $Q$ is strictly positive.
For $M$ small (and fixed charge $e > 0$),
the corresponding eigenvalues $E^0(\theta)$ are given by
perturbative expansion in $M$:
$$ E^0(\theta) = e/2 - \exp (-{1 \over 4e}) \;
M \cos(\theta - \thm) + O(M^2)   \eqno(8) $$
so that, for $M > 0$ the absolute minimum is attained for
$\theta = \thm$.
\endproof

\msk
$\Omega _{\theta \theta _{F}}$ admits a unique extension to the
Weyl algebra in two degrees of freedom, $\AQP \times \Aqp$,
which is a Fock state on $\AQP$ and a Zak state
\citaref{J. Zak, Phys.  Rev. {\bf 168}, 686 (1968);
R. Beaume, J. Manuean, A. Pellet and M. Sirugue, Comm.  Math.  Phys.
{\bf 38}, 29 (1974).}
on $\Aqp$.

The chiral transformations are implementable in
$\piz$ (with $p$ as generator), but spontaneously broken in
each irreducible representation $\Hth$ of $\Aobs$.
In each irreducible representation of the field algebra
$\cal A$ ($\theta _{F}$ fixed)
the gauge transformations are spontaneously broken,
leaving unbroken the discrete subgroup
given by $p,A \mapsto p+n , A+n/e $, $n \in \interi $, which is
implemented by $\exp inq$.
In the analogy with the Schwinger
model such unbroken group corresponds to the
\lq\lq large gauge transformations\rq\rq\ [1--3],[11].
The algebra $\cal A$ can be seen as generated by ${\cal A}_{obs}$
and the \lq\lq charged fields\rq\rq\ $\exp i \alpha p$,
$\alpha \in \reali$,
which play the r\^ole of
$\exp i\alpha  Q_5 $ in the Schwinger model. Clearly, for
mass $M \neq 0 $ there are no $n$--vacua.

In view of the construction of a functional integral representation
of the model, some remarks are necessary on
the irreducible (regular) representations of $\A$
in $\Ldue$; it is easy to see that they are all
unitarily equivalent to those in which
$p$ acts as $-i \partial / \partial \vp + \thf $,
with periodic boundary conditions;
this form is the most convenient one for the control of the
functional integral, since
$\thf$ does not appear in domain problems, and the kernel of
$\exp -tH$, is a periodic function of the angle $\vp$.
For each fixed $\thf $, the representation of $\A$ in $\Ldue$
is also unitarily equivalent to that in $\LdueQth $, with
$\exp i \beta p $ acting as
$$ e^{i \beta p} \phi (Q,\theta) =
\phi (Q, \theta + \beta \ {\rm mod} \ 2\pi) \;
e^{i \beta \thf} \eqno(9) $$
The unitary equivalence is given by
$$ \psi(\vp ,A) = 1/\sqrt{2\pi} \suminf \intdp
  e^{i (A - \thf) (\vp - \theta - 2 \pi n)} \;
\phi (\vp - \theta - 2 \pi n, \theta) \; d\theta
\equiv \intdp \psith (\vp ,A) \, d\theta\eqno(10) $$
where $\phi$ is the wave function in the $Q,q$ representation.
Eq.(10) corresponds to the integral decomposition
$$\Ldue  =  \intdp d\theta \; \Hth  \eqno(11)  $$
over the spectrum of $q$.
In the following for simplicity we will put $e=1$.

\bsk\bsk\goodbreak\ni
{\bf 3. Path integral formulation. Boundary conditions, $\theta$
sectors, winding numbers.}

\bsk
One of the main interests of the model is that it can be
used as a laboratory for
ideas and extrapolations to the functional integral
approach to gauge QFT.
In this perspective one is confronted essentially with two
possible strategies.
The first one exploits the possibility of writing the
Hamiltonian in terms of gauge
invariant fields;
one then restricts the attention
to the algebra of observables
${\cal A}_{obs}$, trivializes  its centre
(by fixing $\exp {iq} = \exp i\theta$), writes
a functional integral representation for the kernel of
$\exp {-tH}$,
and performs the infinite volume limit of the
euclidean correlation functions of
a maximal abelian subalgebra of ${\cal A}_{obs}$;
this is equivalent to the use of a functional
measure defined on trajectories taking values in the spectrum
of $\Aobs$ (see also [8]).
In this way one constructs all
the representations defined by the $\theta$--vacua, for all $\theta$.

This is essentially the strategy followed by [2], where
the free paremeter $\theta$ which appears in the fermion
bosonization in $1+1$ dimensions, and enters in the construction
of the electric field from the fermion currents, plays the r\^ole
of the variable which describes the spectrum of the centre of $\Aobs$.

Similar results are obtained in the functional integral
approach to $QCD$ [1--3], where the $\theta$ vacua
are obtained by summing over the topological number $\nu$
with weight $\exp i \nu \theta$. However, in contrast with
the case of classical trajectories, the topological
classification has no meaning in infinite volume, and therefore
a volume \lq\lq cut--off\rq\rq\ is necessary, and boundary
conditions must be specified (this is also the
r\^ole of the formulation on the sphere, with regularity
at the point at infinity playing the r\^ole of
(special) boundary conditions).
In this approach, the choice of the $\theta$ sector,
and its relation with the sum over topological numbers,
 relies on a particular choice of boundary conditions
\citaref{G. Morchio and F. Strocchi, in {\it Proc. of the
V Int. Conf. on Selected Topics in QFT and  Math. Physics\/},
Liblice June 1989, J. Niederle and J. Fischer eds.,
World Scientific 1990;
Schladming Lectures 1990, in {\it
Fields and Particles}, H. Mitter and W. Schweiger eds.,
Springer--Verlag 1990 pp.171--214.
A different proposal of a dynamical solution of the strong
$CP$ problem has been presented in G. Schierholz,
Nucl. Phys. B (Proc. Suppl.) {\bf 37A}, 203 (1994) },
and it is really done {\it before} the infinite volume limit;
$\theta$ plays therefore the r\^ole of an external, and free,
kinematical constraint, which is independent of
any dynamical (energy density) effect, also
in the infinite volume limit.
This strategy can be realized in the present model, but
crucially requires a {\it special} choice of boundary conditions
(see below).

In our opinion it is of interest to explore a
strategy which does not rely
on low--dimensional pecularities, approaches the construction
of the model  by functional integrating in finite volume
over the field variables which enter
the Lagrangean with {\it generic} boundary conditions,
and then takes the infinite volume limit.
This alternative gives explicit information on
the relevance of boundary conditions and of
free energy density effects, which are expected
to play an important r\^ole also in the (standard)
four--dimensional functional integral
\citaref{See e.g.  L.D. Fadeev and A.A. Slavnov, {\it Gauge Fields.
Introduction to Quantum Theory\/}, 2nd ed., Addison--Wesley 1980.}
over the gauge
potentials $A_{\mu}(x)$ and the Fermion field $\Psi (x)$,
$x\in \reali^{4}$.

\bsk\goodbreak\ni
{\bf Theorem 2}.

\ni
i) {\it For any finite interval $[-T,T]$, the formula
$$ d\tilde\mu_{\xm , \Am , \xp , \Ap , T} \, \xA  = $$
$$ =  \dmz_{\xm , \xp , T} \, \xt \; \,
  \dmz_{\Am , \Ap , T} \, \At \; \,
   e^{-i \intT \dot x A \, d\tau} \;
   e^{M \intT \cos(x - \thm) \, d\tau} \eqno(12) $$
with $\dmz_{\xi_{-}, \xi_{+} , T} \, (\xi(\tau)) $
the conditional Wiener measure on paths
starting at $\xi_{-}$ at $\tau = -T$ and ending at
 $\xi_{+}$ at $\tau = T$, defines a complex measure,
with finite total variation,
absolutely continuous
with respect to the free measure $\dmz(x(\tau)) \;
\dmz(A(\tau))$, and therefore
supported on trajectories $x(\tau) \in \reali$,
$A(\tau) \in \reali$, which are H\"older continuous of
all orders $\alpha < 1/2$.
$d\tilde\mu$ defines a complex measure $d\mu$
on trajectories $\vp(\tau) \in S^1 \equiv [0, 2 \pi)$,
$A(\tau) \in \reali$ by the equation
$$ d\mu_{\vpm , \Am , \vpp , \Ap , T} \, \vpA  =
 \suminf  d\tilde\mu_{\vpm , \Am , \vpp - 2 \pi n , \Ap , T} \, \xA
   \eqno(13) $$
with $\vp (\tau) = x(\tau)$ mod $2\pi$}.

\ni
ii) {\it The measure
$$ d\mu^{\thf}_{\vpm , \Am , \vpp , \Ap , T} \, \vpA  \equiv $$
$$  \suminf e^{i \thf (\vpp - 2\pi n - \vpm)} \;
d\tilde\mu_{\vpm , \Am , \vpp - 2 \pi n , \Ap , T} \, \xA
   \eqno(14) $$
coincides with the measure defined, with boundary conditions
$\vpm, \Am , \vpp , \Ap $, by the kernel
$ \KfvpA $ of $\exp - \tau H$ in $\Ldue$, where $H$ is defined
by eq.(1) with $p = -i \partial / \partial \vp +
\thf$ and periodic boundary conditions in $\vp$:
$$ \KfvpA  =
 \undp \suminf e^{i\vp (A - \thf)} \;
e^{-i(\vp '- 2\pi n)  (A'- \thf)} $$
$$  \intinf   e^{-iq (A-A')} \;
             R_{\tau} (\varphi -q, \varphi ' - q - 2\pi n , q) \, dq
    \eqno(15) $$
where for any fixed $q \in \reali$,
$R(Q,Q',q)$
is the (positive) kernel corresponding to the Hamiltonian
$H_q$ (eq.(4))}.

\ni
iii) {\it Under gauge transformations one has
$$ d\mu_{\vpm , \Am , \vpp , \Ap , T} \;
(\vp(\tau) , A(\tau) - \lambda) = $$
$$ = e^{i [ \lambda ] (\vpp - \vpm)} \;
 d\mu^{(\lambda \; {\rm mod} \; 1)}_{\vpm ,
  \Am + \lambda, \vpp , \Ap + \lambda , T}
  \; (\vp(\tau) , A(\tau))  \eqno(16) $$
$$ K^{\thf = 0}_\tau (\vp, A - \lambda , \vp ' , A' - \lambda)
 = e^{i [ \lambda ] (\vpp - \vpm)} \;
  K^{\thf = \lambda \ {\rm mod} \ 1}_\tau
  (\vp, A , \vp ' , A' )  \eqno(17)  $$
with $[ \lambda ]$ the integer part of $\lambda$, i.e.
$\lambda = [ \lambda ] + (\lambda$ mod $1)$}.

\ni
iv) {\it $K_\tau^\thf$ defines a bounded (hermitean)
semigroup in $\Ldue $; it is irreducible in the sense
that (for any fixed $\tau > 0$) it does not leave
stable any non--trivial subspace of the form
$L^2(B, d\vp \, dA)$, $B \subset
[0,2\pi) \times \reali $}.

\bsk\goodbreak\ni
{\bf Proof}.
We briefly sketch the proof, for more details see [9].
i) One has  to control the convergence of the
discretizations of the integral
$  \exp {-i \intT \dot x A \, d\tau} $
on trajectories in the support of $\dmz \xt \times \dmz \At$,
to a result which is independent of the discretization
and defines a measurable bounded function of the trajectories [9]
(it is sufficient  to consider the case $\xm =\Am =0$).
Since also $   \exp {M \intT \cos (x - \thm) \, d\tau} $
is a measurable bounded function, so is their product.
The absolute continuity of $d\tilde\mu $ with respect to the Wiener
measure and the boundedness of its variation then follow.

\ni
ii) The kernel of $\ittH$ in $\LdueQq $ is
 $  R_{\tau} (Q,Q',q) \, \delta (q'-q) $, with $R_{\tau}$
the kernel of $\exp - \tau H_q$ in $L^2 (\reali)$, $H_q$ being
given by eq.(4). $K_\tau^\thf$ is easily obtained
from this kernel by
using the unitary transformation in eq.(10).
It remains to prove that, for all $\tau > 0$,
$$ \KfvpAdue = \int
   d\mu^{\thf}_{\vp, A , \vp' , A' , \tau} \,
  (\varphi(\tau'), A(\tau'))   \eqno(18) $$
Now, in the r.h.s. of eq.(18),
 the integral in $\dmz \At$ is gaussian and the result is
$$  \undp \suminf e^{i\vp (A - \thf)} \; e^{-i(\vp '- 2\pi n)(A'- \thf)}
  \intinf dq \;  e^{-iq (A-A')} $$
$$   [   \int \dmz_{\vp - q , \vp' - q - 2 \pi n, \tau} \, (x(t))
    e^{- \unm \intt  x^2 (t) \, dt} \;
   e^{M \intt \cos(x (t) - q) \, dt }   ] \eqno(19) $$
The term in square brackets in eq.(19)
is exactly the Feynman--Kac representation
of the kernel $ R_{2 \tau} (\vp-q , \vp'-q'-2\pi n , q)$.

\ni
iii) Eqs. (16) and (17) are proved by explicit calculation.

\ni
iv) is equivalent to the property that $\KfvpA$ is different
from zero almost everywhere. In fact, the integral in the
r.h.s. of eq.(15) defines an entire function of $A$,
since it is the Fourier transform in $q$ of $R_\tau$, which
decreases faster than any exponential. The sum over $n$ does
not destroy analyticity since it
converges uniformily in bounded complex domains,  by the above
decrease properties of $R_\tau$; it follows that
the zeros of $\KfvpA$ are isolated in $A$, and therefore a set of
zero measure with respect to $d\vp \, dA$.

\bsk\goodbreak\ni
{\bf Remark}. Alternatively, given $\KfvpA$, eq.(15), one
could construct the complex measure $\dm^\thf \vpA$ starting
from the cylinder measure defined by the above kernel (with given
boundary values $\vpm , \Am , \vpp , \Ap$);
the existence of a measure (complex, of bounded total variation)
then follows  for $M=0$ from the estimate
$$\int | \KfvpA | \, d\vp' \, dA' \leq  e^{\tau/2}$$
which is proved by explicit computation; for $M \neq 0$ the
result follows from Trotter's product formula, which also controls the
measurability of the mass term with respect to the $M=0$ measure.

\bsk
We now discuss the r\^ole of the boundary conditions and the
winding number interpretation of the functional integral on
observable variables.

Most of the standard wisdom [1-3] on the functional integral approach to
gauge field theory makes use of geometrical ideas which are in general
formulated in infinite volume (in analogy with the classical case).  As
we will also see below, the possibility of incorporating such
geometrical structures in a functional integral theoretical setting is
problematic in infinite volume, since the relevant
measures are not supported on trajectories with definite (classical)
behaviour at infinity.  In our opinion, it is therefore essential to
restrict the use of such structures to finite volume functional
integrals, and to discuss
i) the r\^{o}le of the boundary conditions
and  ii) the stability of topological structures in the
thermodynamical limit.

We recall that, in the $(QED)_{1+1}$ model, the usual analysis, which is
actually done in the case of zero fermion mass and then extrapolated to
the massive case, relies on

\ni
1) the use of (locally
regular) euclidean field configurations which are a pure gauge in a
neighbourhood of euclidean infinity and regular there,
so that they are
classified by a topological number $\nu\in\interi $; at sufficiently
large positive and negative times $\tau_{+}$, $\tau_{-}$,
they are classified
by {\it winding numbers} $n_+$, $n_-$ $(\nu \equiv n_+ -n_-)$;

\ni
2) the (formal)
existence of (time independent) unitary operators ${\cal T}$ which
implement the shift $n \to n+1$ and commute with the
observables ({\it large gauge transformations}).

On the basis of 1) and 2) the Hilbert space is decomposed in sectors
${\cal H}_\theta$, stable under $\A_{obs}$,
which diagonalize ${\cal T}$. In
terms of functional integrals, the $\theta$ sector are formally obtained
by adding to the Lagrangean the topological term
$\theta \, \varepsilon_{\mu\nu} F^{\mu\nu}$.

The same analysis could be applied to $(QED)_{0+1}$ with the volume
replaced by the euclidean time, the regularity at infinity amounting to
$A(\pm \tau) \to A(\pm \infty) $, and
$\exp 2\pi i A(\infty) = \exp 2 \pi i A(-\infty) $, i.e.
$ A(\infty) - A(-\infty) = \nu $; $\nu$
is then the topological invariant of the trajectories
$\exp 2 \pi i A(\tau) :
  \realipunto \to S^1 $, $ \realipunto $ being the compactified
euclidean time.
Without loss of generality one can fix
$\exp 2 \pi i A(\infty) = 1 $,
so that for large positive and negative times
$A(\pm T) \simeq n_\pm$, which play the r\^ole of the winding numbers.
The topological term is $i \theta \int (dA / d\tau) \, d\tau $.

Now, one may show [9] that
the infinite volume measure is a complex cylinder measure
with infinite total variation, and therefore the set
$S_{reg}$ of configurations which are regular at infinity
is not even measurable, since the cylinders are defined
by the values of the variables lying in finite time intervals.
On the other hand, given any extension of the family of measurable
sets which includes $S_{reg}$, the integral of any product of
field variables vanishes when restricted to $S_{reg}$
(as a consequence of time translation invariance and cluster
property), whenever the measure of $S_{reg}$ can be
expressed in terms of countable additivity on cylinder sets
from which $S_{reg}$ can be constructed:
$$  S_{reg} = \cap_n  \cup_{T_1} \cap_{T_2}
    S_{reg} (T_1 , T_2 , 1/n)  $$
$$  S_{reg} (T_1 , T_2 , \varepsilon )  \equiv
   \{ (\vp(\tau) , A(\tau)) : \; | A(\tau_1) - A(\tau_2) |
   < \varepsilon  \ \
   \forall \, \tau_1 , \tau_2 \in [T_1 , T_2]  \} $$
$$ \mu (S_{reg}) = \lim_{n \to \infty} \lim_{T_1 \to \infty}
    \lim_{T_2 \to \infty} \mu ( S_{reg} (T_1 , T_2 , 1/n ) ) $$

We shall actually see that the whole standard picture
($n$ vacua, $\theta$ vacua, functional
integral decomposition etc.) is problematic in infinite volume,
especially in the presence of a fermion mass term.
For these reasons,
in agreement with the well--estabilished wisdom from Statistical
Mechanics, we will first discuss the crucial r\^{o}le of the boundary
conditions for the functional integral in finite volume, and then
perform the infinite volume (\lq\lq thermodynamical\rq\rq ) limit.

The first important feature is that the finite volume euclidean
correlation functions, which satisfy the Osterwalder-Schrader (OS)
positivity
\citaref{K. Osterwalder, in {\it Constructive Quantum Field
Theory}, G. Velo and A.S. Wightman eds., Lecture Notes in Physics vol.
25, Spinger--Verlag 1973.},
give rise [9] to a (finite volume) quantum mechanical
Hilbert Space $\cal H_{OS}$,
which carries an irreducible representation of
the field algebra $\A$ but in general a reducible
representation of the observable algebra $\Aobs$.
The reducibility of $\Aobs$ can be traced
back to the action of the large gauge transformations on the functional
integral and it is the identification of the
\lq\lq irreducible components\rq\rq\ of
the functional integral in finite volume that gives rise to the
$\theta$-vacua picture.

The interpretation which emphasizes the
r\^{o}le of gauge invariance as the relevant requirement leading to the
$\theta$-states is somewhat misleading in our opinion.
In fact, quite generally, {\it every} functional measure
defined on a field algebra $\F$ with (non--trivial) gauge
transformations defines a gauge invariant measure when restricted
to the observable (gauge invariant) subalgebra $\Fobs$, and two
functional measures on $\F$ are physically
equivalent iff they give rise to the same (gauge invariant)
measure on $\Fobs$.
Since any equivalence class always contains,
by general arguments, a measure invariant under
gauge transformations on $\F$, the choice of a measure
within a fixed class is always equivalent to
the choice of a gauge invariant measure, and
corresponds in fact to the so called \lq\lq choice of
gauge fixing\rq\rq .
This applies in particular to the choice of boundary conditions,
which does not affect the locality and gauge invariance properties
of the lagrangean density.
Moreover, the gauge invariance of a measure do not imply the
invariance of the boundary conditions under gauge transformations
{\it defined on boundary variables}, when such boundary
transformations do {\it not} arise from gauge transformations
defined on the whole volume. This is the case of transformations
which change the winding number of the field configurations,
here $A(T) \to A(T) + n_+$,
$A(-T) \to A(-T) + n_-$, with $n_+ \neq n_-$.

To formalize the above statements we first need a notion of
reducibility of the functional measure in finite volume (where we
cannot exploit time translation invariance
\citaref{J. Fr\"ohlich,  Ann. Phys. {\bf 97}, 1 (1976)}.
Since the euclidean
algebra is commutative, reducibility must make reference to the
reconstruction of Quantum Mechanics (QM) and in fact it can be
formulated by exploiting OS positivity: a functional measure $d\mu_T$
satisfying OS positivity will be
said to be QM reducible if it is a convex
combination of functional measures which are OS positive.
It is not difficult to see that the
reducibility of the measure in the above sense is equivalent to the
reducibility of the algebra $\A_E$ generated by the time zero
algebra and by $\exp  -\tau H$, $\tau\geq 0$, in the Hilbert space
$\H_{OS}$ reconstructed from the finite volume correlation
functions through OS positivity.

To discuss the relation between QM reduction and boundary
condition, we start with some general preliminary results;
we denote by $X$ a (topological) space of configuration at fixed
euclidean time, and by $C^0(X)$ the space of
continuous functions on $X$, with the interpretation of the
(abelian) field algebra at fixed time.

\bsk\ni
{\bf Theorem 3.} [9] {\it Let
$d\mu_{\xi_-,\xi_+,T}(\xi(\tau))$, $\xi(\tau)\in X$ be a (complex)
functional measure defined by a hermitian kernel $K_{\tau}(\xi,\xi')$,
which defines a bounded semigroup in $L^2(d\xi)$ (with generator $H$)
irreducible in the sense of Theorem 2, iv); let
$$ d\mu_{\sigma T} \equiv \int
d\sigma(\xi_-,\xi_+) \; d\mu_{\xi_-,\xi_+,T} \, (\xi(\tau))  \eqno(20) $$
where $d\sigma$ is a (boundary) measure of the form
$$  d\sigma(\xi_-,\xi_+) = s(\xi_-,\xi_+) d\xi_- \, d\xi_+ \ \ \ ,
  \ \ \ \ \ \  s(\xi_-,\xi_+) \in L^2(d\xi_-,d\xi_+)    $$
Then the correlation functions defined by the
functional measure $d\mu_{\sigma  T}$ satisfy the OS positivity iff
$s(\xi_-,\xi_+)$ is of the form
$$ s(\xi_-,\xi_+)=  \sum_{i}
\lambda_i \, \bar\psi_i(\xi_-) \, \psi_i(\xi_+) \ \ , \ \ \
\psi_i \in L^2, \eqno(21)  $$
$d\mu_{\sigma T}$ is then of the form
$\sum\lambda_i \, d\mu_{\psi_i , T}$,
where each $d\mu_{\psi_i ,T}$ satisfies
the OS positivity.
The OS reconstructed space $\cal H_{OS}$ is a direct sum
of copies ${\cal H}_i \simeq L^2(X, d\xi)$, obtained by OS
reconstruction with boundary conditions given by $\psi_i$; each
${\cal H}_i$ is stable under the algebra $\A_E$ generated by
$ C^0 (X)$ and $\exp - \tau H$, $\tau > 0$;
$\A_E$ is irreducible in ${\cal H}_i$, as a consequence of
the irreducibility of $K_{\tau}$}.

Thus, for irreducible boundary conditions, i.e.
$s(\xi_-,\xi_+)=\bar{\psi}(\xi_-) \, \psi(\xi_+)$, and only
in this case, the integral in the
euclidean variables with measure (20) represents the euclidean
correlation functions,
on the state $\psi$ with wave function $\psi(\xi)$,
of a QM model living in $L^2(d\xi)$ with Hamiltonian $H$:
$$ (\psi  , e^{-H(T+\tau_1)} \, F_1(\xi) \, e^{-H(\tau_2-\tau_1)} \dots
F_n \, e^{-H(T-\tau_n)} \, \psi) = $$
$$  \int \bar\psi(\xi_-) \, \psi(\xi_+) \,
d\mu_{\xi_-,\xi_+,T} \, (\xi(\tau)) \,d\xi_- d\xi_+  \eqno(22) $$

The next issue is the decomposition of $ {\cal H}_{OS} $ into irreducible
representations of an \lq\lq observable subalgebra\rq\rq .
In the above general framework, we denote by
$C^0_{obs}(X)$ a subalgebra of $C^0(X)$, which will be interpreted
as the observable algebra at fixed (euclidean) time.

\ni {\bf Theorem 4} [9] {\it
Under the assumptions of Theorem 3,
with irreducible $d\mu_{\sigma  T}$, let
$U$ be a (non-trivial) operator acting on $C^0(X)$
with the properties:

\ni
1) $U: C^0(X) \rightarrow C^0(X)$

\ni
2) $ [ U,  K_{\tau}] = 0 $

\ni
3) $U$ commutes with $C^0_{obs}(X)$, as multiplication operators.

\ni
Then the restriction of the measure $d\mu_{\sigma  T}$ to
the algebra generated by $\prod_\tau C^0_{obs}(X) $
is QM reducible, and the algebra $\A_{E,obs}$ generated by
$C^0_{obs}(X)$ and  $\exp - \tau H $, $\tau > 0$
is reducible in ${\cal H}_{OS}$. If $U$ is normal
(as an operator in $\H_{OS}$), then
its spectral projectors reduce the representation of $\A_{E,obs}$,
and the corresponding Hilbert space decomposition is obtained
by decomposing the boundary wave function $\psi (\xi)$
according to the spectrum of $U$}.

In the $(QED)_{0+1}$ model the r\^{o}le of $U$ is played
by the unitary operator $\exp i q$ implementing the
\lq\lq large gauge transformations\rq\rq , and
the corresponding spectral decomposition of the boundary conditions
gives the $ \theta$ decomposition of
$\H_{OS}$.
Such decomposition is trivial in the representation of the functional
integral in terms of the variables $E, q$, since in this case $U$ is a
multiplication operator in the euclidean variables.
In the ($\varphi , A $)
representation the decomposition is less trivial,
and it is given by eq.(10).
In fact,
the decomposition (10), applied to the wave functions which define
the boundary conditions, gives the $\theta$ decomposition of the
functional integral  of variables in the {\it observable} euclidean
algebra, generated by the gauge invariant exponentials
$\exp i(2\pi m A(\tau)+n\varphi(\tau))$, $m,n \in \interi$,
with an obvous winding number interpretation (for simplicity,
here and in the following,
we put $\thf = 0 $):

$$  \int_{- \infty}^{\infty} \, dA_- \, dA_+ \, \int_{0}^{2\pi}
d\varphi_- \, d\varphi_+ \, \bar\psi (\vpm , \Am ) \,
\psi (\vpp , \Ap)  $$
$$  \int  d\mu_{\vpm , \Am , \vpp , \Ap ,T} \,
(\varphi(\tau), A(\tau))  \   F_1 (\varphi(\tau_1), A(\tau_1)) \ldots
F_n(\varphi(\tau_n), A(\tau_n)) = $$
$$  = \undp
   \int_{0}^{1} dB_- \, dB_+ \,
\int_{0}^{2\pi} d\varphi_- \, d\varphi_+ \,
\int_{0}^{2\pi} d\theta_- \, d\theta_+
\sum_{n , \nu , k_- , k_+  \in \interi}
e^{-i(B_- - \theta_F) (\vpm  - \theta_-  - 2\pi k_-)} $$
$$  e^{i(B_+ - \theta_F) (\vpp - \theta_+  - 2\pi k_+)}
e^{-in (\theta_+ - \theta_-)}
 e^{i\nu(\vpp - \theta_+ - 2\pi k_+)}
  \; F_1 \ldots F_n $$
$$ \bar\phi (\vpm -\theta_- -2\pi k_- , \theta_-) \,
       \phi (\vpp -\theta_+ -2\pi k_+ , \theta_+)
  d\mu_{ \vpm , B_-, \vpp , B_+ + \nu, T} \,
(\vp(\tau) , A(\tau) - n) \eqno(23) $$
where we have put
$A_\pm = B_\pm + n_\pm$, $B_\pm \in [0,1)$, $n_\pm \in \interi $,
$n_+ = n_- + \nu \equiv n + \nu$, and used eq.(16).
Since $F_1,\ldots ,F_n$ are
invariant under $A \to A+n$, we can drop  $n$ in the argument of
the measure, so that the sum on $n$ gives a
$\delta(\theta_- - \theta_+)$
and the integration over $\theta_+$, with a relabelling
$\theta_- \to \theta$, yields a decomposition of the form
$$ \int_{0}^{2\pi} d\theta \, \sum_{\nu\in\interi}
e^{-i\nu\theta} \, d\mu^\thf_{\vpm , B_-, \vpp, B_+, \nu , T} \,
(\varphi(\tau),B(\tau)) $$
$$  \int dB_- \, dB_+ \, d\vpm  \, d\vpp \,
\bar\psi_\theta(\vpm , B_- )
\psi_\theta ( \vpp , B_+ )  \   F_1 \ldots F_n \eqno(24) $$
with
$d\mu_{\vpm , B_-, \vpp , B_+ \nu , T} \,
 (\vp(\tau) , B(\tau))$ the restriction of
$d\mu_{\vpm , B_-, \vpp , B_+ +\nu , T} \, (\vp(\tau) ,A(\tau))$
to the euclidean observable algebra,
with $B(\tau) = A(\tau)$ mod $1$.

A sharp isolation of a $\theta$ state representation requires
to choose, as boundary condition, a wave function
$\psi_\theta (\vp ,A)$ in the space  of the representation $\pi_0$
described in Theorem 1,
$$ \psi_\theta(\vp ,A) = \sum_{n}
e^{i(A - \theta_F) (\varphi - \theta -2\pi n)} \,
\Phi(\varphi - \theta - 2\pi n)   \eqno(25) $$
 with $\Phi\in L^2$; one must then
integrate with respect to
$d\varphi_- \, d\varphi_+ \, dA_+$ and take the {\it ergodic
mean} in $A_-$.  This shows that the construction of $\theta$
states (in finite volume) requires boundary conditions which are
crucially \lq\lq non--local\rq\rq\ in the variable $A$.
The construction is
very close to that of Bloch $\theta$ sectors
for a particle in a periodic potential [8].

It is worthwhile to stress that the above decomposition into pure phases
of $\A_{E,obs}$ is done in finite volume and it is
radically different from
the usual phase decompositions, which are
obtained in Statistical Mechanics
and (euclidean) Field Theory
{\it after} the infinite volume limit [16].
This makes the discussion of symmetry breaking very
different from the standard case, since the r\^{o}le of the
thermodynamical limit is substantially different.

The above decomposition of the functional integral in finite volume has
a winding number interpretation.  The integer $\nu$ is the winding
number which classifies the trajectories of $A$ as trajectories on the
spectrum of $\exp 2\pi i A$, which is a circle,
exactly as in the case of a
particle in a periodic potential $W(x)$, with the periodic functions
of $x$ playing the r\^{o}le of the observables [8].
In $QED_{0+1}$ such
topological structure crucially depends on the compactness of the chiral
group, since otherwise no non--trivial functions of $A$ would be
observable.
As in the case of periodic potentials, it is easy to see that
the topological numbers of the configurations do not remain
bounded in the infinite volume limit, and therefore
the discussion in finite volume
is essential for the use of the topological classification
of the configurations.

We may now easily discuss the effect of the addition of a
(gauge invariant) \lq\lq topological term\rq\rq\
to the Lagrangean, namely a term
$\theta_L \dot A$, leading to an interaction of the form
$$ i \theta_L \int_{-T}^{T} \dot{A}(\tau) \, d\tau =
i \theta_L (A(T)-A(-T)) \eqno(26) $$
in the euclidean action.
{}From the above discussion it follows that the effect on the measure
$d\mu(\vp(\tau), A(\tau))$ is a change of boundary conditions
$$  d\sigma(\vpm , A_-, \vpp , A_+) \rightarrow
e^{i\theta_L (A_+ - A_-)} \,
d\sigma (\vpm , A_-, \vpp , A_+) \eqno(27)$$
This means that the addition of the topological term
amounts to a change of boundary conditions
$\psi(\vp , A) \to \exp (i \theta_L A) \, \psi (\vp , A)$;
in particular, for boundary conditions (25) leading to a
$\theta$ state, the effect of the term (26) is to yield
the sector labelled by $\theta + \theta_L$.

Thus, the standars link between the lagrangean parameter
$\theta_L$ and the parameter $\theta$, which labels the
irreducible representations of the observable algebra,
only holds if the boundary conditions are those of eq.(25),
with $\theta = 0$.
For {\it generic boundary conditions}, which define reducible
representations of the observable algebra,
the addition of the topological term only shifts the support
of the $\theta$ reduction; as we will see, for $M \neq 0$,
the infinite volume limit {\it removes}  such a reducibility and
selects (generically in the boundary conditions)
the sector with $\theta = \thm$,
{\it independently of the topological term}.

The above considerations do not apply to the formulation
based on a functional integral  on the spectrum of the
euclidean {\it observable} algebra; there, the addition
of the topological term (l.h.s. of eq.(26)) does not
reduce to a change of boundary conditions because
$\exp i\theta_L A(\pm T)$ is not an observable, and eq.(26)
is not available.
Therefore, in that strategy, the construction of $\theta$ sectors
is not related to the decomposition of the boundary
conditions; in fact
the angle $\theta$ enters directly in the measure (through the
the kernel of $\exp -\tau H$) as a free parameter.

\bsk\bsk\ni
{\bf 4. Thermodynamical limit.  Convergence to the ground state with
$\theta=\theta_M$ in the massive case.}

As usual, the construction of the lowest energy state(s) requires the
control of the thermodynamical limit $(T\rightarrow\infty)$ of
properly normalized correlation functions, eq.(22).
Unlike the standard
case (of models with strictly positive kernels,
with positive boundary conditions), here the so constructed state
depends in general on the boundary conditions (b.c.).
We concentrate our
discussion on the limit of the correlation functions of observables.
The main problem is whether the reduction into sectors,
which appears in
finite volume for generic b.c., survives the $T\rightarrow\infty$
limit.  For the case $M\neq 0$, the situation is very similar to that of
particle in a periodic potential, since the lowest energy
state is for small
$M$ unique (Theorem 1); this implies (even if the ground state
does not belong to the space of the OS reconstruction in finite volume,
see Theorem 1) that
for generic b.c. the correlation functions converge to the
expectations on the  unique ground state, which belongs
to the sector ${\cal H}_{\theta=\theta_M}$
({\it no $CP$ violation}).

\bsk\ni
{\bf Theorem 5}. {\it $\forall f(\varphi,A) \in(L^1 \cap
L^2)([0,2\pi) \times \reali , d\varphi \, dA)$,
such that
$$ f_\theta^{0} \equiv
( \psi_\theta^{0}  , f )_{L^2 ([0,2\pi) \times \reali ,
d\varphi \, dA)}    \eqno(28) $$
with $\psi_{\theta}^{0}(\varphi ,A)$ the wave function of the lowest
energy state in ${\cal H}_\theta$, does not vanish for $\theta =
\theta_M $, and for all observables
$F_1 \ldots F_n$,
$F_k = \exp i(2\pi m_kA(\tau_k) + l_k\varphi(\tau_k))$  one has
$$  \lim_{T\to\infty}     Z(f,T)^{-1}
  \intinf dA_- \, dA_+ \int_{0}^{2\pi} d\varphi_- \, d\varphi_+
  \bar{f}(\varphi_- , A_-) \, f(\varphi_+ , A_+) $$
$$ \int d\mu_{\varphi_-, A_- , \varphi_+ , A_+ , T} \,
(\varphi(\tau), A(\tau))  \   F_1\ldots F_n  =  $$
$$ =(\psi_{\theta_M}^{0} , e^{i(2\pi m_1 A + l_1\varphi} \,
e^{- (\tau_2-\tau_1) (H-E^0(\theta_M)) } \ldots
e^{i(2\pi m_n A + l_n\varphi)} \, \psi_{\theta_M}^{0})  \eqno(29)  $$
for $M$ small.
For $M=0$, the above limit gives
$$  \int_{0}^{2\pi} d\theta \,
g(\theta) \, (\psi_{\theta}^{0} , e^{i(2\pi m_1 A + l_1 \varphi)} \,
e^{-(\tau_2-\tau_1) (H-E^0(\theta)) } \,  \ldots  \,
e^{i(2\pi m_nA + l_n\varphi)}\psi_{\theta}^{0}) \eqno(30)  $$
with $ g(\theta) \equiv
|f_{\theta}^{0}|^2 / \int_{0}^{2\pi}|f_q^{0}|^2 \, dq$.
The normalization constant $Z(f,T)$ is given by the functional integral
for $F_k=1$,\ $k=1,\ldots,n$}.

\msk\ni
{\bf Proof.}\ The proof is
similar to that for a particle in a periodic potential [8]
(for more details see [9]). The essential ingredients are:
i) the Feynman--Kac representation, eq.(22),
 ii) the finite volume decomposition of the (OS) Hilbert space into
$\theta$ sectors,  iii) the discretness of the spectrum of $H$ in
each sector ${\cal H}_\theta$,  iv) the continuity in $\theta$ of
$|f_{\theta}^{0}|$,
which gives the weight of the component of $f_\theta (\varphi ,A)$,
eq.(10), over the ground state
$\psi_{\theta}^{0} (\varphi ,A)$,  v) the uniqueness of the ground
state $\psi_{\theta_M}^{0}$ (Theorem 1), and the non--zero gap
$\inf_{\theta, n > 0}(E^n (\theta) - E^0 (\theta)) \equiv \delta > 0$,
for $M\neq 0$ and small.
In fact, the l.h.s. of eq. (29) is of the form
$$ [ \int_{0}^{2\pi} d\theta \;
|f_{\theta}^{0}|^2 \; (\psi_{\theta}^{0} ,
e^{-(T+\tau_1)H} \, e^{i(2\pi m_1 A + l_1\varphi)} \,  \ldots  \,
e^{-(T-\tau_n)H} \,
\psi_{\theta}^{0}) +  O(e^{-2(E^0 (\theta) + \delta) T}) ] $$
$$   [ \int_{0}^{2\pi} d\theta \;
|f_{\theta}^{0}|^2 \;
(\psi_{\theta}^{0}, e^{-2TH} \psi_{\theta}^{0}) +
O(e^{-2(E^0 (\theta) + \delta)T}) ]^{-1}. \eqno(31) $$
For large $T$, since $|f_{\theta}^{0}|$ is continuous and
 $|f_{\theta_M}^{0}|\neq 0$ only the first terms in the
square brakets survive.
Moreover,
$$ (\psi_{\theta}^{0}, e^{-(T-\tau_1)H} \,
e^{i(2\pi m_1A + l_1\varphi)} \,  \ldots \,
e^{i(2\pi m_nA + l_n\varphi)} \, e^{-(T-\tau_n)H} \,
\psi_{\theta}^{0}) = e^{-2 E^0 (\theta) T} G(\theta) \eqno(32) $$
with $G(\theta)$ the correlation function with a
renormalized Hamiltonian
$H \rightarrow  H - E^0(\theta)$;
since $E^0(\theta) > E^0(\theta_M)$,
$\forall \theta \neq \theta_M$,
$$ |f_{\theta}^{0}|^2 \, e^{-2E^0 (\theta) T}
 [ \int _{0}^{2\pi} d\theta \, |f_{\theta}^{0}|^2
e^{-2E^0 (\theta) T} ]^{-1}
\to \delta (\theta-\theta_M) \eqno(33) $$
for $T \to \infty$ in the sense of measures,
and eq.(29) follows since $G(\theta)$ is continuous in $\theta$.

For $M=0$, the result follows immediately from the fact
that $E^0(\theta) $ is independent of $\theta$, so that
the r.h.s. of eq.(32) becomes
$ \exp -2 E^{0} T  \, G(\theta)$,
and the l.h.s. of eq.(33) is independent of $T$,
and actually given by $g(\theta)$.
(The normalization factor $Z(f,T)$ is strictly positive,
since it is given by the l.h.s. of eq.(22), with $F_i = 1$, and
$\exp (- \tau H) > 0 $).

The above proof makes clear the substantial
difference between the $M=0$
and the $M\neq 0$ cases and shows that the limit $M \to 0$ does
not commute with the thermodynamical limit $(T \to \infty)$ [9][13].
Since the essential ingredient is the existence (for $M\neq 0$) of a
unique absolute minimum of $E^0(\theta)$, the above result applies
also to higher dimensions on the basis of a
$\theta$ sector decompositions
of the form of eqs.(24),(31) with the free energy density
in the $\theta$ sector playing
the r\^{o}le of $E^0 (\theta)$.

In contrast with the approach which restricts the attention
to the algebra of observables, the strategy discussed above,
which integrates over lagrangean field variables, allows to
compare the free energy density for different values
of $\theta$, and to exploit the fact that it has a (unique) minimum,
at $\theta = \thm$.

The implications on the strong $CP$ problem is that in the first case
the renormalization is done {\it after} having fixed the
value of $\theta$ (which therefore appears as a free parameter),
whereas in the second case the renormalization automatically
preserves the property
that the free energy density has a unique minimum,
at $\theta = \thm$,
which holds generically in the values of the parameters,
so that a \lq\lq reduction\rq\rq\
automatically take place in the infinite volume limit,
with the selection of $\theta = \thm$
({\it natural strong $CP$ symmetry}).

\vfill\eject\immediate\closeout\fileref
                \par\vfill\eject
                \null\msk
                \centerline{\bf References}
                \bsk
                \input ref.tmp

\bye